# Wang-Landau algorithm: A theoretical analysis of the saturation of the error

R. E. Belardinelli[a)] and V. D. Pereyra[b)]
*Departamento de Física, Laboratorio de Ciencias de Superficie, Universidad Nacional de San Luis, CONICET, Chacabuco 917, 5700 San Luis, Argentina*

In this work we present a theoretical analysis of the convergence of the Wang-Landau algorithm [Phys. Rev. Lett. **86**, 2050 (2001)] which was introduced years ago to calculate the density of states in statistical models. We study the dynamical behavior of the error in the calculation of the density of states. We conclude that the source of the saturation of the error is due to the decreasing variations of the refinement parameter. To overcome this limitation, we present an analytical treatment in which the refinement parameter is scaled down as a power law instead of exponentially. An extension of the analysis to the *N*-fold way variation of the method is also discussed.

## I. INTRODUCTION

Monte Carlo methods are well suited for the simulation of large many-body problems, particularly for the study of phase transitions and critical phenomena.[1,2]

Many conventional Monte Carlo algorithms such as Metropolis importance sampling,[3] Swendsen-Wang cluster flipping,[4] etc., generate a canonical distribution at a given temperature. Such distributions are so narrow that, with conventional Monte Carlo simulations, multiple runs are required to determine thermodynamic quantities over a significant range of temperatures.

For a second-order phase transition in unfrustrated systems, the problem of critical slowing down was solved by a cluster update algorithm.[4] For first-order phase transitions and in systems with many free energy minima such as frustrated magnets or spin glasses, a similar problem of long tunneling times between local minima arises.

Several methods were developed to overcome this problem, including the multicanonical method,[5–14] and related entropic sampling method,[15] broad histogram method,[16–18] multibonding simulations,[19] etc.

The multicanonical ensemble method proposed by Berg et al.[5–7] estimates the density of states (DOS) $g(E)$ (the number of all possible states or configurations for an energy level $E$ of system) first, then performs a random walk with a flat histogram in the desired region in the phase space. This method has been proven as very efficient in studying the first-order phase transitions, where simple canonical simulations have difficulty in overcoming the tunneling barrier between coexisting phases at the transition temperature.[5,8–14]

The entropic sampling method, which is basically equivalent to multicanonical ensemble sampling,[15] is in iterative process used to calculate the microcanonical entropy defined as $S(E)=\ln[g(E)]$. The method was applied to the two-dimensional (2D) ten-sate ($Q=10$) Potts model and the three-dimensional (3D) Ising model.[15]

The broad histogram method introduced by de Oliveira et al.[16–18] calculates the density of states by estimating the probabilities of possible transitions between all possible states of a random walk in energy space. However, the method presents systematic errors even for simple models such as the Ising model at small system sizes.[18] In fact, those methods based on accumulation of histogram entries have the problem of scalability for large systems, suffering from systematic errors when systems are large.[20]

Wang and Landau[20] introduced a new and efficient algorithm using a random walk in energy space to obtain an estimation of the density of states for statistical models. The method has been successfully used in many problems of statistical physics, biophysics, and others.[20–29] The method is based on independent random walks which are performed over adjacent overlapped energy regions, providing the density of states. In that way, thermodynamic observables, including the free energy over a wide range of temperature, can be calculated with one single simulation.

Since Wang and Landau introduced the multiple-range random walk algorithm to calculate the density of states, there have been numerous improvements proposed[29–38] and studies of the efficiency and convergence of this method.[32,37,38] Particularly, Zhou and Batt[32] have presented a mathematical analysis of the Wang-Landau (WL) algorithm. They give a proof of the convergence and the sources of errors for the WL algorithm and the strategies for improvement. The main conclusions of their work are (i) the density of state is encoded in the average histogram; (ii) the fluctuations of the histogram, proportional to $1/\sqrt{\ln f}$, cause statistical errors, which can be reduced by averaging over multiple $g(E)$; (iii) the correlation between adjacent records in the histogram introduces a systematic error, which is reduced at small $f$.

Earl and Deem in Ref. 38 have derived a general condi-

---

[a)] rbelar@unsl.edu.ar
[b)] vpereyra@unsl.edu.ar

tion for the convergence of a Monte Carlo method whose history dependence is contained within the simulated density distribution. The authors concluded that (i) the detail balance needs only be satisfied asymptotically and (ii) the calculated density of states approaches to the real one with an error proportional to $1/t$.

On the other hand, several limitations of the method remain still unsolved, as, for example, the behavior of the tunneling time which is a bound for the performance of any flat-histogram algorithm, as is discussed in Ref. 36, where it is shown that it limits the convergence in the WL algorithm.

Berg have proposed a multicanonical simulation scheme where the first part is the WL algorithm and the second part is a standard Markov chain Monte Carlo simulation for which convergence is proven.[39]

Berg and Janke have introduced a cluster version of the Wang-Landau algorithm together with a subsequent multibondic simulation. This method improves the efficiency of the conventional WL or multicanonical approach by power laws in the lattice size for systems such as 2D and 3D Ising models.[19]

Although several authors say that the WL algorithm converge,[32,38] one of the limitations of the method and the subsequent improved algorithms based on it is the saturation in the error.

The saturation in the error and therefore the nonconvergence of the WL method were also suggested by Landau and co-workers already in Refs. 37 and 40. Particularly the improvement of the $N$-fold way variation introduced by Malakis et al.[41] does not avoid the saturation in the error.

In fact, those methods in which the refinement parameter vary lower faster than $1/t$ (with $t$ the Monte Carlo time) determine that the calculated density of states reaches a constant value for long times, therefore the error saturates. To overcome this limitation, we recently introduced a modified algorithm in which the refinement parameter is scaled down as $1/t$ instead of exponentially.[42] This new algorithm allows the calculation of the density of states faster and more accurately than with the original WL algorithm due to the fact that the calculated density of state function approaches asymptotically the exact value.

In this work we present a theoretical demonstration of the saturation in the error in the WL algorithm and deduce analytically the optimum choice of the refinement parameter $F$ as $F(t)=1/t$, without using the histogram flatness condition as was introduced in Ref. 42. The resulting algorithm is more efficient than the original WL method and other variations like those obtained by using $N$-fold way algorithm.[30,41] Due that the exact density of state is, in general, rarely known, we implement all the calculations in a two-dimensional $L \times L$ Ising model, where $g(E)$ is exactly known for size up to $L=50$.[43]

The outline of the paper is as follows. In Sec. II, we introduce the time behavior of the Wang-Landau algorithm, demonstrating the origin of the saturation of the error. In Sec. III, we give the analytical bases of the $1/t$ algorithm introduced previously in Ref. 42. In Sec. IV, we discuss the dynamical behavior of the $N$-fold way version of the WL algorithm, introducing the corresponding $1/t$ modification of our algorithm using such method. Finally, in Sec. V, we discuss the results and give our conclusions.

## II. THEORETICAL ANALYSIS OF THE CONVERGENCE OF THE WANG-LANDAU ALGORITHM

In the original Wang-Landau algorithm,[20] an initial energy range of interest is identified, $E_{\min} \leq E \leq E_{\max}$, and a random walk is performed in this range. During the random walk, two histograms are updated: one for the density of states $g(E)$, which represents the current or running estimate, and one for number of visits to distinct energy states $H(E)$. Before the simulation begins, $H(E)$ is set to zero and $g(E)$ is set to unity, both uniformly. The random walk is performed by choosing an initial state $i$ in the energy range $E_{\min} \leq E \leq E_{\max}$. Trial moves are then attempted and moves are accepted according to the transition probability

$$p(E_i \rightarrow E_f) = \min\left[1, \frac{g(E_i)}{g(E_f)}\right], \quad (1)$$

where $E_i(E_f)$ are the initial (final) states, respectively, and $p(E_i \rightarrow E_f)$ is the transition probability from the energy level $E_i$ to $E_f$. Whenever a trial move is accepted, a histogram entry corresponding to state $n$ is incremented according to $H(E_n) = H(E_n) + 1$ and $g(E_n) = g(E_n) \times f$, where $f$ is convergence factor, that is, generally initialized as $e$. If a move is rejected, the new configuration is discarded and the histogram entry corresponding to the old configuration is incremented according to $H(E_o) = H(E_o) + 1$; at the same time the density of states is incremented according to $g(E_o) = g(E_o) \times f$. This process is repeated until the energy histogram becomes sufficiently flat. When that happens, the energy histogram $H$ is reset to zero everywhere and the convergence factor is decreased, usually according to $f_{k+1} = \sqrt{f_k}$, where $f_k$ is the convergence factor corresponding to stage $k$. The process is continued until $f$ becomes sufficiently close to 1 [say, $f < \exp(10^{-8})$].

To analyze the dynamical behavior of the WL algorithm, it is necessary to define different quantities. The histogram $H'(E,t)$ is defined in an analogous way as the histogram $H$ used in the WL algorithm, but is not reset in the whole simulation. Its mean value after time $t$ is defined as

$$\langle H'(t) \rangle = \frac{1}{N} \sum_E H'(E,t), \quad (2)$$

where $N$ is the number of states of different energies and $H'(E,t)$ stands for the mean height of the histogram in $E$ at time $t$ (with $t=j/N$, where $j$ is the number of trial moves attempted). The time $t$ is related to the numbers of accessible energy states. For example, in a two-dimensional Ising model the time $t$ is given by $t=j/(L^2-1)$, where $L$ is the size of the system (if we consider only one energy window to calculate the DOS).

Instead of the density of states it is more convenient to define $S(E,t) = \ln(g(E,t))$, usually named entropy. This definition is generally used in order to fit all possible values of $g(E)$ into double precision numbers. Its mean value is defined as

$$\langle S(t) \rangle = \frac{1}{N} \sum_E S(E,t). \quad (3)$$

The error $\epsilon(E,t)$ is defined as

$$\epsilon(E,t) = \left| 1 - \frac{\ln[g_n(E,t)]}{\ln[g_{ex}(E)]} \right| = \left| \frac{S_{ex}(E) - S_n(E,t)}{S_{ex}(E)} \right| \quad (4)$$

and its mean value as

$$\langle \epsilon(t) \rangle = \frac{1}{(N-1)} \sum_E \epsilon(E,t). \quad (5)$$

$g_n(E,t)$ is normalized with respect to the exact DOS at the ground state, that is,

$$g_n(E,t) = \frac{g(E,t) g_{ex}(E_G)}{g(E_G,t)}, \quad (6)$$

where $g(E,t)$, $g_{ex}(E_G)$, and $g(E_G,t)$ are the simulated value, the exact values at the ground state, and the simulated value at the ground state of the DOS, respectively. Therefore,

$$S_n(E,t) = \ln[g_n(E,t)] = S(E,t) - S(E_G,t) + S_{ex}(E_G). \quad (7)$$

It is also convenient to define the function $F = \ln[f]$, then $F_{k+1} = F_k/m$ (with $m > 1$).

Note that $F_k$ is always positive and monotonically decreasing.

It is necessary to emphasize that Eq. (6) breaks down for any classical continuous system, for which the entropy diverges both at very low and very high energies, therefore other normalization schemes must be introduced in order to avoid the divergence of the entropy. For instance, choosing other reference state with energy $E'$ in such a way that $g_{ex}(E') \neq 0$.

Then $S(E,t)$ can be written as

$$S(E,t) = \sum_{i=1}^{t} [H'(E,i) - H'(E,i-1)] F_{i-1}, \quad (8)$$

where $F_0 = \text{const}$ [usually, $F_0 = \log(e) \simeq 0.434\,294\,4$].

The flatness condition of the histogram $H(E,t)$ and Eq. (1) guarantee that $F_{i+1}$ takes the value $F_i$ a finite number of times before eventually decreasing to $F_i/m$. Moreover $H'(E,i) - H'(E,i-1)$ is finite. Since $\sum_{k=1}^{\infty} F_k$ is convergent, the series $S(E,t)$ is also convergent for any value of $m$.

Let us rewrite the error given in Eq. (4) as

$$\epsilon(E,t) = \frac{|\Delta S(E,t) - \Delta S_{ex}(E,t)|}{S_{ex}(E)}, \quad (9)$$

where $\Delta S(E,t) = S(E,t) - S(E_G,t)$ and $\Delta S_{ex}(E) = S_{ex}(E) - S_{ex}(E_G)$. Since $S(E,t)$ and $S(E_G,t)$ are convergent, $\Delta S(E,t) \to \text{const}$ when $t \to \infty$. However, due to the stochastic character of the Markov process, both quantities are not statistically equal, $\Delta S(E,t) \neq \Delta S_{ex}(E)$, then the mean value of the error goes to a constant limit as time increases to infinity.

Let us show, as an example, the calculation of the error for the 2D Ising model at the boundaries of the energy range, $E = 2L^2$. Note that the ground state energy is $E = -2L^2$ for this model. It is well known that $S_{ex}(-2L^2) = S_{ex}(-2L^2) = \ln(2)$, hence

$$\epsilon(2L^2,t) = \left| \frac{S(2L^2,t) - S(-2L^2,t)}{S_{ex}(2L^2)} \right|. \quad (10)$$

Then, using Eq. (8) and the argument of convergence, one can conclude that $S(2L^2,t) \to K(2L^2)$ and $S(-2L^2,t) \to K(-2L^2)$ when $t \to \infty$, where $K(2L^2)$ and $K(-2L^2)$ are constant. Due to the stochastic character of the process, these quantities are not statistically equal, therefore $\epsilon(2L^2,t)$ saturates, and the mean value of the error $\langle \epsilon \rangle$ saturates as well. Similar arguments can be used to demonstrate the saturation of the error for each energy level. In the present case of the 2D Ising model, the exact value of the density of states $g_{ex}(E)$ was obtained from the method developed in Ref. 43.

In conclusion, the WL algorithm does not converge to the true value of the DOS because the series given in Eq. (8) is convergent.

## III. AVOIDING THE SATURATION OF THE ERROR: THE $1/t$ ALGORITHM

In order to avoid the saturation of the error, the series given in Eq. (8) must be made divergent, with $F_k$ monotonically decreasing and $F_k$ tending to zero as fast as possible. The last condition is not arbitrary since it guarantees that $g(E,t) \to g_{ex}(E)$ rapidly, reducing the computational time.

The series which fulfills such conditions is the harmonic series $\sum_{k=0}^{\infty} 1/k^p$ with $P \leq 1$, the fast converging being the one with $p=1$.

In order to make the WL algorithm more efficient, the refinement parameter must take, for long times, the following time functionality:

$$F(t) = \frac{1}{t}, \quad (11)$$

which is updated at each Monte Carlo step.

Initially, $F_0 = 1$, then at short time, and as was proposed previously,[42] we choose $F(t) > 1/t$ (i.e., $F_{i+1} = F_i/2$ as in the original WL algorithm) in such a way that the algorithm visits all the energy configurations as fast as possible. As soon as $F_i \leq 1/t$, that is at time $t_c$, the algorithm changes and the refinement parameter takes the time functionality described above. Then, using Eq. (8), the entropy takes the form

$$S(E,t) = \sum_{k=1}^{t_c-1} [H'(E,k) - H'(E,k-1)] F_k$$
$$+ \sum_{k=t_c}^{t} [H'(E,k) - H'(E,k-1)] \frac{1}{k}, \quad (12)$$

for $t \leq t_c$, $F_{k+1} = F_k/2$, or $F_{k+1} = F_k$,

From the definition of Eq. (3), and replacing the expression of $S(E,t)$ given in Eq. (12), we obtain

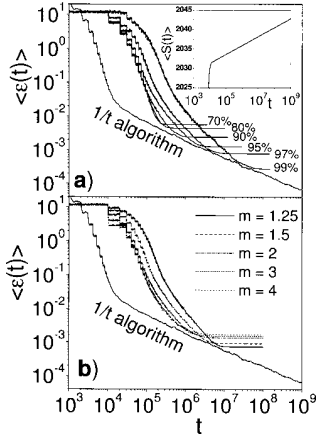

FIG. 1. (a) Comparison between the error $\langle\epsilon(t)\rangle$ calculated using the original WL method with different flatness conditions and the error obtained using our $1/t$ algorithm, for a two-dimensional Ising model with nearest-neighbor lateral interaction in squares lattice with $L=8$. In the inset $\langle S(t)\rangle$ is shown vs time $t$, in semilog plot, to emphasize its logarithmic behavior. (b) Behavior of the error as a function of time, for different values of the parameter $m$ for a flatness condition of 95%.

$$\langle S(t)\rangle = \frac{1}{N}\sum_E \sum_{k=1}^{t_c-1}[H'(E,k)-H'(E,k-1)]F_k$$
$$+\sum_{k=t_c}^{t}[H'(E,k)-H'(E,k-1)]\frac{1}{k}. \quad (13)$$

We can rearrange the last equation, summing first on the energy range, then considering that on average $\langle H'(t)-H'(t-1)\rangle \approx 1$, to finally obtain

$$\langle S(t)\rangle \propto \ln(t). \quad (14)$$

This quantity is necessary to show the dynamics of the algorithm and not to obtain the observable which are calculated using the expression of the entropy given in Eq. (12).

In Fig. 1(a), we show a comparison between the error $\langle\epsilon(t)\rangle$ calculated using the WL method with different flatness conditions and the error obtained using our algorithm, for the Ising model in a 2D squares lattice with $L=8$ and $F_{k+1}=F_k/m$ (with $m=2$). In the inset, the behavior of the $\langle S(t)\rangle$ is shown in semilog plot for our algorithm. As one can see, after the initial time $t_c$ the function $\langle S\rangle$ takes the form given in Eq. (14). In Fig. 1(b) we show the behavior of the error as a function of time, for different values of the parameter $m$ and flatness condition of 95%. The accuracy of the WL method decreases as $m$ increases.

Note that the error, in our algorithm, is proportional to $1/\sqrt{t}$ instead of $1/t$, as is predicted by Earl and Deem in Ref. 38.

Note that instead of Eq. (11), we can propose a more general time dependence for the refinement parameter as

$$F(t)=\frac{c}{t^p}. \quad (15)$$

However, a simple observation of the behavior of $F(t)$ as a function of $p$ and $c$ and the corresponding effect on the error $\langle\epsilon(t)\rangle$ show that the best choice to optimize the efficiency of the algorithm is that for $p=1$ and $c=1$. In fact, in Fig. 2(a) we show the function $F(t)$ as a function of $t$ for different values of the exponent $p$ with $c=1$. The effect on the error $\langle\epsilon(t)\rangle$ is shown in Fig. 2(b). For $p\leq 1$, the error goes as $\langle\epsilon\rangle \propto \sqrt{F(t)}$ for long time (asymptotic regime), one also observes that the value which optimizes the error is $p=1$. As is discussed above for $p>1$, the error saturates. In Figs. 3(a) and 3(b) we show the function $F(t)$ and the corresponding error $\langle\epsilon(t)\rangle$, as a function of $t$ for $p=1$ and three different values of $c$ ($c=0.1,1,10$). The error goes as $\langle\epsilon(t)\rangle \propto \sqrt{c/t}$ for $c\geq 1$ and changes the time behavior for $c<1$. In fact, the error follows a power law dependence with an exponent bigger than $-0.5$. However, in all cases the error does not saturate.

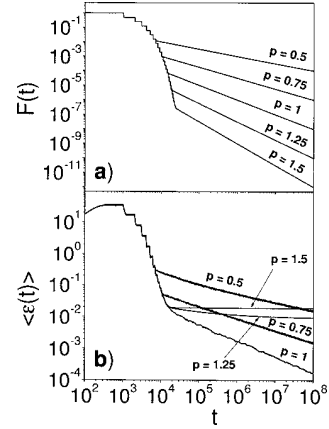

FIG. 2. (a) Dynamical behavior of $F(t)$ for different values of the exponent $p$ and $c=1$. (b) Error $\langle\epsilon(t)\rangle$ vs time $t$, for the same values of $p$ and $c=1$ [see Eq. (15)].

In Fig. 4 we show the error $\langle\epsilon(t)\rangle$ versus the refinement parameter $F$ calculated using the original WL algorithm, for different flatness conditions.

Note that the behavior of the error follows the law $\sqrt{F}$ before the error saturates, confirming the conclusion (ii) predicted in Ref. 32. On the other hand, no matter how large the flatness condition is, the error saturates in all cases for smaller values of $F$, demonstrating that the calculated DOS does not converge to the exact value; this fact is in contradiction with the third conclusion predicted by Zhou and Batt in Ref. 32.

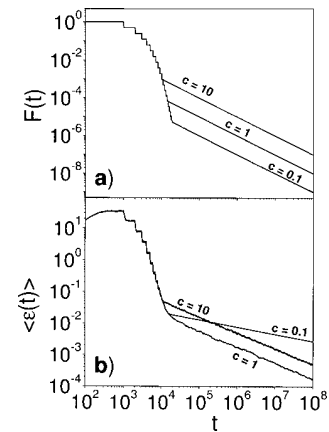

FIG. 3. (a) Dynamical behavior of $F(t)$ for different values of $c$ and $p=1$. (b) Error $\langle\epsilon(t)\rangle$ vs time $t$, for the same values of $c$ and $p=1$ [see Eq. (15)].

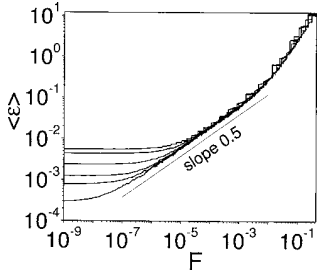

FIG. 4. Error $\langle\epsilon(t)\rangle$ vs the refinement parameter, calculated using the original WL method, for different values of the flatness condition (from top to bottom, 70%, 80%, 90%, 95%, 97%, and 99%).

## IV. ON THE 1/t ALGORITHM USING THE N-FOLD WAY

In this section we discuss the dynamical behavior of the N-fold version of the WL algorithm.[30,41] in order to show that also in this case the error saturates.

To overcome such a problem we introduce a new version of our method, the N-fold way 1/t algorithm. As in original version of the WL N-fold way algorithm,[30] we assume a spin system which may be in state $\sigma$ with energy $E \in I = [E_{min}, E_{max}]$, whereby $I$ denotes the energy range for which we wish to estimate $g(E)$. All spins are then partitioned into classes according to their energetic local environment, i.e., the energy difference $\Delta E_i$ a spin flip will cause. For the special case of the two-dimensional nearest-neighbor Ising model, each spin belongs to one of only $M=10$ classes. The total probability $P$ of any spin of class $i$ being flipped is given by

$$P(\Delta E_i) = n(\sigma, \Delta E_i) p(E \to E + \Delta E_i), \quad i=1,\ldots,M, \quad (16)$$

with $n(\sigma, \Delta E_i)$ being the number of spins of state $\sigma$ which belong to class $i$ and $p(E \to E + \Delta E_i)$ being given by

$$p(E \to E + \Delta E_i) = \begin{cases} \min(1, g(E)/g(E+\Delta E_i)), & \text{if } E+\Delta E_i \in I \\ 0, & \text{if } E+\Delta E_i \notin I. \end{cases} \quad (17)$$

To determine the class from which to flip a spin, one calculates the numbers

$$Q_m = \sum_{i \leq m} P(\Delta E_i), \quad m=1,\ldots,M \quad \text{and} \quad Q_0=0, \quad (18)$$

which are the integrated probabilities for a spin flip within the first $m$ classes. Hence a class is selected by generating a random number RND such that $0 < \text{RND} \leq Q_M$, and taking class $m$ if $Q_{m-1} < \text{RND} \leq Q_m$. The spin to be flipped is chosen from this class with equal probabilities. The numbers $n(\sigma, \Delta E_i)$ change upon flipping the spin. Due to the flip, the spin and its interacting neighbors will change classes and correspondingly the numbers $n(\sigma, \Delta E_i)$ will differ from their predecessors. Finally, one has to determine the average lifetime $\tau$ of the resulting state, i.e., one has to find out how many times the move just made would be rejected on average in the usual update scheme. The probability that the first random number would produce a flip is $\hat{P} = Q_M/L^2$. Therefore, one has for the probability that exactly $n$ random numbers will result in a new configuration,

$$\bar{P}_n = \hat{P}(1-\hat{P})^{n-1}. \quad (19)$$

Thus the average lifetime becomes

$$\tau = \sum_{n=1}^{\infty} n\bar{P}_n = \sum_{n=1}^{\infty} n\hat{P}(1-\hat{P})^{n-1} = \frac{L^2}{Q_M}, \quad (20)$$

Based on these definitions, we can describe the N-fold way version of the 1/t-algorithm as follows.

(i) Choose an initial configuration and set $H(E)=0$, $S(E)=0$ for all $E$: $t=0$, $F_0=\log(e) \simeq 0.434\,294\,4$ and also fix $F_{\text{final}}$.
(ii) Determine (update) the probabilities $p(E \to E + \Delta E_i)$ and the $Q_m$'s of the (initial) configuration using Eqs. (16)–(18).
(iii) Determine the average lifetime $\tau$ of (initial) configuration via Eq. (20).
(iv) Increment histogram, density of states, the time $t$, and update $f_i$,

$$H(E) \to H(E) + 1,$$

$$S(E) \to S(E) + \Delta S(E),$$

$$t \to t + \tau/N,$$

$$f_i \to f_{i+1},$$

with

$$\Delta S(E) = \begin{cases} F_i \tau & \text{if } F_i \tau \leq F_0 \\ F_0 & \text{if } F_i \tau > F_0, \end{cases} \quad (21)$$

and

$$F_{i+1} = \begin{cases} F_i & \text{if } F_i \tau \leq F_0 \\ \Delta S(E)/\tau & \text{if } F_i \tau > F_0. \end{cases} \quad (22)$$

In the case where $F_i/F_{i+1} > 2$, we set

$$F_{i+1} \to F_i/2. \quad (23)$$

(v) After some fixed sweeps check $H(E)$; if $H(E) \neq 0 \,\forall E$ then refine $F_i$ according to $F_{i+1}=F_i/2$ and $H(E)$ is reset.
(vi) If $F_{i+1} \leq 1/t$ then $F_{i+1}=F=1/t$. Then $H(E)$ and Eqs. (22) and (23) are not used in the rest of the simulation, while from Eq. (21) we only use $\Delta S(E)=F_i\tau$.
(vii) If $F < F_{\text{final}}$ the simulation stops.
(viii) Determine the $Q_m$'s [the $p(E \to E + \Delta E_i)$'s are not updated here].
(ix) Choose and flip spin as described above.
(x) Go to (ii).

The initial value of the refinement parameter $F_0$ has a crucial importance at the beginning of the algorithm and can determine the subsequent dynamics. In fact, we observe that choosing $F_0$ small, the algorithm wastes a long time visiting the most accessible states. On the contrary, if $F_0$ is large enough, the algorithm wastes a long time visiting the least accessible states. The more efficient election is thus $F_0 \approx 0.5$.

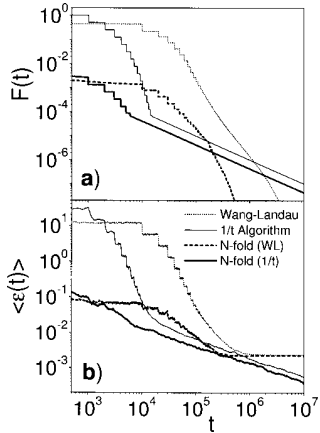

FIG. 5. (a) Refinement parameter $F(t)$ vs time and (b) dynamical behavior of the error for the four methods mentioned in the text.

As in the algorithm introduced in Ref. 30, $\Delta S(E)$ is kept below or equal $F_0$, avoiding that $Q_M = 0$ which would terminate the iteration procedure described above. The time $t$ is the accumulate average lifetime used as a refinement parameter in our algorithm, similarly to $1/t$ in the algorithm introduced previously in Ref. 42.

In Fig. 5(b), we have plotted the dynamical behavior of the average error $\langle \varepsilon(t) \rangle$ as a function of the Monte Carlo time, both the original and the $N$-fold way version of the WL algorithm saturate, while our $1/t$ version and the corresponding $N$-fold way variation do not. A comparison of these curves demonstrates that our methods are in both cases more efficient than the original algorithms. In Fig. 5(a), we have plotted the corresponding refinement parameter for each algorithm. As one observes, $1/t$ behaves as limiting curves, as soon as the refinement parameter for the WL algorithm in both the original and the $N$-fold way versions is lower than this limiting curve, the corresponding errors saturate.

Clearly, the error saturates also in the $N$-fold way variation of the WL algorithm. The reason of such behavior was explained in Sec. II of the present paper. In fact, on the average $F_i \tau_i > F_{i+1} \tau_{i+1}$ and due to that $S(E,t)$ can be written as a series, with a kernel $F_i \tau$, then the series that it represents this function is convergent. Using the same argument, to avoid the saturation in the error, we have to choose the refinement parameter as $F = 1/t$.

## V. DISCUSSION AND CONCLUSIONS

In this work, we discussed one of the weaknesses as of the well known Wang-Landau algorithm,[20] namely, the saturation of the error or the nonconvergence of calculated density of states to the exact value. We presented an analytical demonstration of the nonconvergence to the exact DOS in the original version of the WL algorithm. We have also shown that the saturation in the error appears not only in the original version of the WL algorithm but in the $N$-fold way variation of such method.[30] Alternatively, we deduced analytically the way to avoid the saturation of the error and gave an adequate form to the refinement parameter. This new algorithm, the so-called $1/t$ algorithm already introduced by us in Ref. 42, was then extended within the $N$-fold way scheme.

The nonconvergence of the original WL algorithm and other previous version, including $N$-fold way method, seemed very difficult to believe. However, in view of our results we are able to discuss some of these statements raised by other authors.[32,38] Particularly, some of the conclusions presented by Zhou and Bhatt[32] are not satisfied completely by the original WL algorithm. In fact, the second conclusion is true before the error saturates, after that it is no longer valid. On the other hand, the correlation between adjacent records in the histogram introduces a systematic error, which is not reduced by small $F$ as is demonstrated in the present paper, therefore the third conclusion is not valid.

It is interesting to emphasize that Landau et al. in Ref. 41 suggest that $\ln(f_{\text{final}})$ cannot be chosen arbitrarily small or the modified $\ln[g(E)]$ will not differ from the unmodified one to within the number of digits in the double precision numbers used in the simulation. If this happens, the algorithm no longer converges to the true value, and the program may run forever. If $\ln(f_{\text{final}})$ within the double precision range is too small, the calculation might take excessively long to finish.

Summarizing, in this work, we have presented an analytical proof of the origin of the error saturation in the WL algorithms and a method to avoid it. We have chosen, as a test laboratory, a discrete system, namely, the Ising model. However, the mathematical arguments of the source of the error for the WL algorithm seem to be more general and can be extended to all algorithms which consider a refinement parameter that change, according to the flatness condition of the energy histogram, with a law that decreases faster than $1/t$. In all these cases, a saturation of the error for the calculation of the density of states can be guaranteed. In fact, due that the entropy $S(E,t)$ can be expressed as a series in which $F(t)$ is the kernel. In those algorithms where $F_k = F_{k-1}/m$ (with any value of $m > 1$), the resulting series converges to a finite value, then the error saturates.

The simplest way to avoid the saturation in the error is to choose a refinement parameter that depends on time as $F(t) = t^{-p}$ with $p \leq 1$ (the optimum choice is $p = 1$). In these cases the series becomes divergent and the calculated density of states approaches asymptotically to the exact values as $\propto t^{-p/2}$. This choice results in a more accurate and more efficient algorithm than other methods.

Finally, we have considered the extension of our algorithm to the $N$-fold way method. It was shown that the adequate refinement parameter is $F = 1/t$, where $t$ is the accumulate average lifetime. This definition ensures that the DOS approaches asymptotically the exact value more efficiently than using the $N$-fold way original WL algorithm which saturates at long times.

## ACKNOWLEDGMENTS

We Thank Professor G. Zgrablich for reading the manuscript. This work is partially supported by the CONICET (Argentina).